\def\year{2021}\relax

\documentclass[letterpaper]{article} 
\usepackage{aaai21}  
\usepackage{times}  
\usepackage{helvet} 
\usepackage{courier}  
\usepackage[hyphens]{url}  
\usepackage{graphicx} 
\usepackage{natbib}  
\usepackage{caption} 
\usepackage{tikz}
\usepackage{tabularx}

\usetikzlibrary{fadings}
\usetikzlibrary{patterns}
\usetikzlibrary{shadows.blur}
\usetikzlibrary{shapes}

\usepackage{amssymb} 
\usepackage{subcaption} 
\usepackage{booktabs} 

\nocopyright
\title{A Survey on Echo Chambers on Social Media: Description, Detection and Mitigation}
\author {
    Faisal Alatawi, 
    Lu Cheng,
    Anique Tahir,
    Mansooreh Karami,
    Bohan Jiang,
    Tyler Black,
    Huan Liu
}
\affiliations {
    Arizona State University - DMML Lab \\
    \{faalataw,lcheng35,artahir,mkarami,bjiang14,tqblack,huanliu\}@asu.edu
}

\begin{document}

\maketitle

\begin{abstract}
    Echo chambers on social media are a significant problem that can elicit a number of negative consequences, most recently affecting the response to COVID-19. Echo chambers promote conspiracy theories about the virus and are found to be linked to vaccine hesitancy, less compliance with mask mandates, and the practice of social distancing. Moreover, the problem of echo chambers is connected to other pertinent issues like political polarization and the spread of misinformation. An \textit{echo chamber} is defined as a network of users in which users only interact with opinions that support their pre-existing beliefs and opinions, and they exclude and discredit other viewpoints. This survey aims to examine the echo chamber phenomenon on social media from a social computing perspective and provide a blueprint for possible solutions. We survey the related literature to understand the attributes of echo chambers and how they affect the individual and society at large. Additionally, we show the mechanisms, both algorithmic and psychological, that lead to the formation of echo chambers. These mechanisms could be manifested in two forms: (1)~the bias of social media’s recommender systems and (2)~internal biases such as confirmation bias and homophily. While it is immensely challenging to mitigate internal biases, there has been great efforts seeking to mitigate the bias of recommender systems. These recommender systems take advantage of our own biases to personalize content recommendations to keep us engaged in order to watch more ads. Therefore, we further investigate different computational approaches for echo chamber detection and prevention, mainly based around recommender systems. 
\end{abstract}

\section{Introduction}
\label{sec:1_introduction}
Having access to verified and trusted information is crucial in the midst of the COVID-19 pandemic, one of the most significant health crises~\cite{Mallah2021COVID} in recent history. Exposure to misinformation on social media has been linked to COVID-19 vaccine hesitancy~\cite{Loomba2021Impact}, the belief that 5G towers spread the virus~\cite{Ahmed20205G}, the misconception that a COVID-19 vaccine candidate caused the death of trial participants\footnote{\url{https://www.politifact.com/factchecks/2020/dec/11/blog-posting/two-vaccine-trial-participants-died-fda-didnt-conn/}}, and a widely held view that the virus is a conspiracy or a bioweapon~\cite{Douglas2021Conspiracy}. These beliefs threaten the response to the pandemic and promote actions that can lead to the spread of the virus. In this regard, the Alan Turing Institute~\cite{Seger2020Tackling} classified \textit{epistemic security} as a fundamental challenge for society when facing a situation that requires taking collective action to respond to crises (e.g., global pandemics) or complex challenges (e.g., climate change). They define \textit{epistemic security} as reliably preventing threats to the production, distribution, consumption, and assessment of reliable information within a society. Echo chambers on social media are identified~\cite{Seger2020Tackling} as one of the core threats to epistemic security as they can drastically increase the spread and even creation of misinformation on social media~\cite{DelVicario2019Polarization,DelVicario2016Spreading,Zollo2017Debunking,Zollo2018Misinformation}. The presence of misinformation on social media is a well-documented problem~\cite{Kai2020Combating,Wu2019Misinformation}. Social media is a prominent source of news and information about COVID-19 and other current events for most of us. Currently, more than half of adults in the US say that they get their news from social media~\cite{Shearer2021News}.

The misinformation percolating in social media echo chambers can have dangerous ramifications in the physical world. For instance, misinformation on social media sites regarding COVID-19 has been proven to be associated with less compliance with social distancing policies~\cite{Bridgman2020Causes}. This behavior and others led to the spread of the virus on a large scale. This is in part because social media provided a climate for this type of misinformation to spread: Social media sites lack the editorial supervision that traditional news media outlets have~\cite{Kai2017Fake}. In addition, the spread of misinformation has been linked to the echo chamber phenomenon on social media~\cite{Tornberg2018Viral,Chiou2018Fake,DelVicario2016Spreading}. Many of the widespread COVID-19 rumors and mass misinformation campaigns, for instance, have been linked to social media echo chambers. As shown in the study conducted on conservative media users by Romer and Jamieson~\cite{Romer2021Patterns}, 61\% of participants believe that the CDC (Centers for Disease Control and Prevention) ``are exaggerating the danger posed by the coronavirus to damage the Trump presidency.'' All of this leads us to conclude that echo chambers are a severe problem that needs to be understood and addressed.


To study echo chambers in a computational perspective, the first step is to have a definition that highlights the core features of echo chambers on social media. This definition should avoid conflating echo chambers with filter bubbles and political polarization. Both are often incorrectly used interchangeably with echo chambers. While political polarization is an attribute of social media echo chambers (as we will explain in section \ref{sec:2_attributes}), the difference between echo chambers and filter bubbles is more subtle. Unfortunately, that is why many echo chamber definitions fail to capture the complete concept of social media echo chambers. The most common definition~\cite{Dubois2018Overstated,Sindermann2020Age,Choi2020Rumor,Eck2021Climate} of echo chambers comes from Jamieson and Cappella~\cite{Jamieson2008Echo}. They define an echo chamber as a metaphor to capture the ways that messages are amplified and reverberate through a media platform. Jamieson and Cappella describe the echo chamber as a ``bounded, enclosed media space that has the potential to both magnify the messages delivered within it and insulate them from rebuttal''~\cite{Jamieson2008Echo}.

To be more specific, we focus on three crucial features unique to echo chambers on social media: (1)~they are a network of users, (2)~the content shared in that network is one-sided and very similar in the stance and opinion on different topics~\cite{Garimella2018Political}, and (3)~outside voices are discredited and actively excluded from the discussion~\cite{Nguyen2020Echo}. Therefore, in this survey, we define an \textit{\textbf{echo chamber}} as a network of users in which users only interact with opinions that support their pre-existing beliefs and opinions, and they exclude and discredit other viewpoints.

This definition helps us to differentiates echo chambers from filter bubbles based on these three aspects~\cite{Nguyen2020Echo}: (1)~the reason behind the lack of access to dissenting opinions, (2)~how outside sources are viewed, and (3)~the effect of exposure to debunking and counterevidence. First, echo-chamber members actively exclude and discredit voices that share dissenting opinions, while in filter bubbles, these voices are left out, most likely unintentionally due to over-personalization~\cite{Pariser2011Filter}. Second, echo-chamber members distrust all outside sources, while users trapped in a filter bubble lack exposure to relevant information and arguments, and once they are exposed to these opinions, they might agree with it. Finally, exposure to counterevidence can break a filter bubble but may actually backfire and reinforce an echo chamber.

\begin{figure*}[t]
\centering
\includegraphics[width=\textwidth]{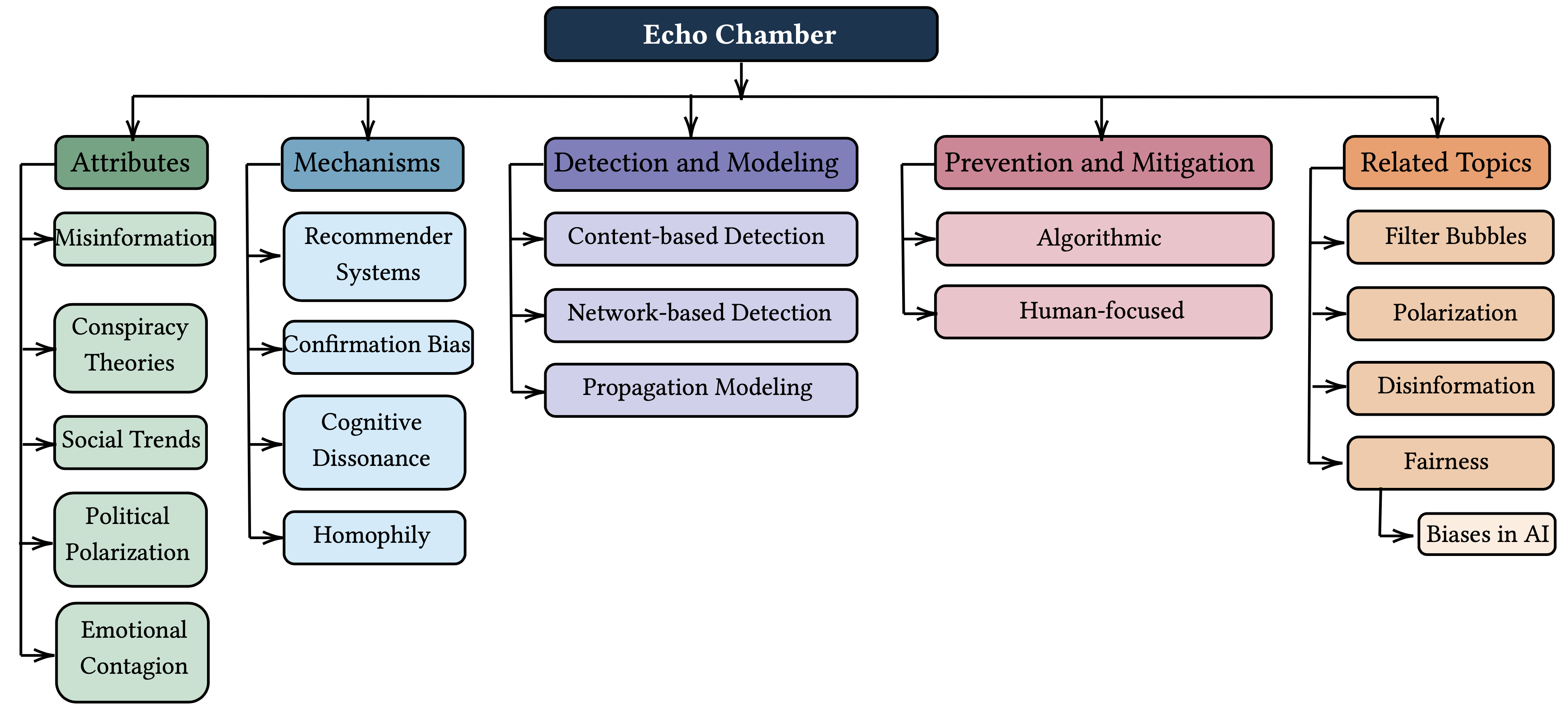}
\caption{An overview of Echo Chamber Research. This figure illustrates the structure of our survey as well as the related topics and concepts to echo chambers.} \label{fig::outline}
\end{figure*}


In this survey, we focus on echo chambers on social media platforms. Social media create an environment where we can communicate with anyone worldwide and share our ideas and opinions about various topics. On the other hand, social media facilitates the spread of mass misinformation and disinformation. The manner in which social media recommends both content and people to follow has helped form filter bubbles and echo chambers that exclude users from being exposed to others' opinions.

The research interest in online echo chambers is primarily due to fear of negative influence from online social media on society. It started with the 2000 US presidential election and Sunstein's 2001 book ``Republic.com''~\cite{Sunstein2001Republic} where he talked about the effect of the internet on group polarization. The second wave coincided with the 2008 US elections of President Obama, where Jamieson and Cappella~\cite{Jamieson2008Echo} published their seminal book ``Echo Chamber.'' This book ignited most of the current work on echo chambers. The third wave happened around 2011 and coincided with the development of AI and machine learning, especially in content personalizations and recommendations. In 2011, Pariser~\cite{Pariser2011Filter} published his book about how social media creates information bubbles, and he coined the term filter bubbles. In 2016, the age of post-truth and mass misinformation started the fourth wave of interest in echo chambers. The focus was on the spread of fake news and rumors. Many research cited echo chambers as a possible explanation for fake news~\cite{Kai2017Fake}. Finally, the fifth and current wave coincided with the rise of new echo chambers like QAnon and the events of the 2020 election in the US and the global Covid-19 pandemic.

Social media recommender systems recommend content that utilizes users' psychological biases like confirmation bias, cognitive dissonance, and homophily. The goal of social media's recommender systems is to recommend content that keeps users engaged with the platform to spend more time watching more ads. We do not criticize the financial motive behind social media. However, we argue that social media has a social responsibility to not cause harm to society by promoting polarized content that leads to the formation of echo chambers.

Although echo chambers are not unique to social media, social media accelerates the formation of them and sustains them. Social media sites have three main features that make them a perfect environment for echo chambers: (1)~there is no geographical limitation to join an echo chamber; (2)~there is no social price to pay if they share fringe beliefs; (3)~no matter how fringe the beliefs are, you most likely will find someone who shares them with you. To illustrate this, consider the ``flat earth'' echo chamber. Before social media, if someone said that they believed that the earth was flat or they questioned the shape of the earth, they would have been mocked and their level of education would have been questioned. When social media is added to the equation, professing belief in the flat earth ``theory'' bears no social cost. Furthermore, you most likely will find someone who shares your belief; especially, in the absence of geographical limitations. After all, there are 7 billion people on earth, certainly some of them believe in fringe ideas. Of course, these fringe ideas existed before social media, but we argue that social media makes them spread faster and on a larger scale. It took 40 years for the Flat Earth Society to reach 3,500\footnote{\url{https://www.nytimes.com/2001/03/25/us/charles-johnson-76-proponent-of-flat-earth.html}} members. In a fraction of that time, their Twitter account\footnote{\url{https://twitter.com/FlatEarthOrg}} has gained 94,000 followers.

Our survey focuses on the social computational perspective of echo chamber research. Our goal is to understand the echo chamber phenomenon, its mechanisms, attributes, and how to detect it, and how to mitigate it, or even better, prevent it from happening in the first place. We hope to provide a blueprint for a solution to the echo chamber problem. Although an individual intervention has a limited effect, we believe that a solution based on modifying the way social media recommend content is a promising direction. This solution depends on the detection of echo chambers. Because without knowing if social media has an echo chamber or not, we cannot mitigate/prevent its effect. Detecting echo chambers can help us understand their interaction with other members and how they grow and form. This information could help us to prevent future echo chambers from forming. Based on this observation, we structured our paper as shown in Figure 1.


\bigskip \noindent \textbf{Intended Audience.} This work is mainly intended for researchers in the fields of social computing, machine learning, data mining, artificial intelligence, and ethical artificial intelligence. Additionally, we hope that this work interests social media providers to overcome the echo chamber problem and its negative effects~(see Section \ref{sec:2_attributes} for more).


\bigskip \noindent \textbf{Differences from Existing Surveys.} The main difference between our survey and others arises from the fact that the field of echo chambers on social media lacks a comprehensive survey and a review that covers the topic from the perspective of social computing. For instance, Nguyen~\cite{Nguyen2020Echo} provides an excellent review of the echo chamber and the filter bubbles phenomena from the perspective of philosophy. Levy and Razin~\cite{Levy2019EchoEconomic} survey the economics literature on echo chambers, and they focus on the mechanisms that create echo chambers. In addition, there are a number of surveys on related topics such as filter bubbles~\cite{Bruns2019Filter}, political polarization~\cite{Tucker2018Social}, and misinformation~\cite{Kai2017Fake,Kai2020Combating,Zhou2018Fake,Sharma2019Combating}. Our work adds to the field of echo chambers by highlighting the work done in social computing and focuses on the possible methods to address the echo chamber problems and other related issues.


 \bigskip \noindent \textbf{The structure of the paper and our contributions}. Figure \ref{fig::outline} shows the structure of our survey and outlines the research interests and topics related to the research on echo chambers on social media. \textit{\textbf{Our contribution}} can be summarized as follows:
 
\begin{itemize}
    
    \item We define the echo chamber phenomenon on social media, and we clarify how it differs from similar social media-related phenomena such as filter bubbles and political polarization (Section \ref{sec:1_introduction}).

    \item We discuss attributes of social media's echo chambers (Section \ref{sec:2_attributes}). Specifically, we focus on the social impacts and potential risks related to echo chambers. We highlight the interaction between echo chambers' members and the society regarding misinformation, conspiracy theories, social trends, political polarization, and emotional contagion.

    \item We list the mechanisms that lead to the formation of echo chambers from the perspectives of recommender systems, human psychology, and biases (Section \ref{sec:3_mechanisms}). Our goal is to explore the mechanisms that cause echo chambers in the first place to understand the echo chamber phenomenon, which is the first step to solve it.

    \item We review the methods and features that could be used to detect echo chambers in social media platforms (Section \ref{sec:4_detection}). Additionally, we examine methods to model echo chambers to study their formation and interaction with people outside the echo chamber. Our goal is to explore and exploit the methods and the features to detect echo chambers, which is crucial for echo chamber prevention.

    \item We discuss methods to prevent echo chambers from forming, and in case they already formed, how to mitigate their negative effect (Section \ref{sec:5_prevention}).

    \item We discuss the open problems related to echo chambers and the challenges that could be encountered while working on them (Section \ref{sec:6_challenges}).

    \item We document some of the datasets that could be used in future work on echo chambers (Appendix \ref{appendix::data}).

\end{itemize}

\section{Attributes of Echo Chambers}
\label{sec:2_attributes}
In this section, we illustrate five common attributes of echo chambers: diffusion of misinformation, spreading of conspiracy theory, creation of social trends, political polarization, and emotional contagion of users. We introduce the common attributes of echo chambers at the very beginning for giving a preliminary insight of what echo chambers look like in online social media~\cite{jiang2021mechanisms}. We also discuss their different outcomes, social impacts, and potential risks.

\subsection{Diffusion of Misinformation}
Misinformation refers to false information that is spread, regardless of intent to mislead\footnote{\url{https://www.dictionary.com/browse/misinformation}}. Nowadays, mainstream social media platforms are used by people due to easy access, low cost, and fast dissemination of news pieces~\cite{Kai2017Fake}. However, the quality and credibility of the content spread in social media is considered lower than traditional news media because of a lack of regulatory authority. Thus, people manipulate the public by leveraging echo chambers to propagate misinformation~\cite{Tornberg2018Viral}. Echo chambers exclude dissenting opinions, make users insist on their confirmation bias, and let misinformation go viral. The effect of echo chambers on the spread of rumors~\cite{Choi2020Rumor} and misinformation~\cite{Tornberg2018Viral,Chou2018Addressing,Cota2019Quantifying} has been proven by many researchers.

Despite early efforts have been undertaken to mitigate online misinformation~\cite{Kai2020Combating, gottlieb2020information, shu2019detecting}, the COVID-19 related misinformation were widely spreading on social media as a global crisis~\cite{li2020mm}. Existing methods have been ineffective for the COVID-19 disinfodemic because: (1)~the contents are novel and highly deceptive; (2)~the dissemination is rapid; and (3)~they require experts with domain knowledge to fact-check. Echo chambers in social media manipulate not only influencers but also common people to become misinformation spreaders. They enable users to intentionally or unintentionally disseminate misinformation faster~\cite{Tornberg2018Viral}. Misinformation spread in echo chambers usually contains three characteristics: (1)~similar misinformation is frequently scrolled and repeated to the users; (2)~the contents are inflammatory and emotional; and (3)~meant to mislead people by exploiting social cognition and cognitive biases. Because the diffusion of misinformation can cause rampant negative effects and is the most common attributes of echo chambers in social media, echo chamber detecting methods should take it into consideration.

\subsection{Spreading of Conspiracy Theories}
A common definition of conspiracy theory is a belief that some covert but influential organization meet in secret agreement with the purpose of attaining some malevolent goal~\cite{bale2007political}. Echo chambers in social media have provided fertile grounds for conspiracy theories to spread quicker~\cite{metaxas2017infamous}. Existing research illustrates that various conspiracy theories have been circulating through mainstream media~\cite{grimes2020health, juhasz2017political, van2019echo}. Conspiracy theories are attempts to explain the ultimate causes of significant social and political events and circumstances~\cite{douglas2019understanding}.
Conspiracy believers use social media to find each other, disseminate conspiracy contents, and share fringe viewpoints. Conspiracy theories express and amplify anxieties and fears about losing control of religious, political, or social order~\cite{marwick2017media}. Unlike misinformation, conspiracy theories are often strongly believed by governments. This results in catastrophic impact to society. For example, AIDS denial by the government of South Africa, was estimated to have resulted in the deaths of 333,000 people~\cite{simelela2015political}.

During the COVID-19 pandemic, despite the fact that many COVID-19 vaccines have been shown to be safe and effective for generating an immune response~\cite{jackson2020mrna}, they need to be accepted by at least 55\% to 85\% of the population to provide herd immunity depending on countries~\cite{kwok2020herd}. However, the COVID-19 vaccine-related conspiracy theories have been widely circulating on online social media. For example, viral social media posts say that Bill Gates intends to implant microchips in people through the coronavirus vaccine~\footnote{\url{https://www.bbc.com/news/52847648}}\textsuperscript{,}\footnote{\url{https://www.usatoday.com/story/news/factcheck/2020/06/12/fact-check-bill-gates-isnt-planning-implant-microchips-via-vaccines/3171405001/}}. Other conspiracy theories such as that the pandemic is a bioweapon~\cite{pennycook2020fighting}, and that the 5G towers and cellphone cause the coronavirus pandemic~\cite{meese2020covid19}. Such information can spread fears through online social media, marking a shift towards declines in societal confidence and trust, and limits public uptake of COVID-19 vaccines.

\subsection{Creation of Social Trends}
Social media platforms present temporal popular topics as social trends on the main website to attract user’s attention. Top trends are usually summarized in several trending words or hashtags. For example, ``\#JohnsonVariant'' was in Twitter's top trending topics as the Britain's Prime Minister Boris Johnson announce to lift most remaining COVID-19 precautions in England on July 19, 2021\footnote{\url{https://www.forbes.com/sites/brucelee/2021/07/09/johnsonvariant-trends-as-boris-johnson-will-lift-covid-19-precautions-despite-delta-variant/?sh=e9de7c662b17}}. People use the ``Johnson Variant'' to refer to the more contagious Delta variant of COVID-19, and express angry, concerns, and depression to the administration. The creation of such trending topics is one of the common attribute of echo chambers in social media.

Many studies have tried to discover the important factors that cause trending topics~\cite{mathioudakis2010twittermonitor, romero2011influence, wu2007novelty}. Asur et al.~\cite{asur2011trends} found that the resonance of the content with the users of the social network is crucial. They further define the measurement of ``resonance'' in three parts: (1)~the novelty of content; (2)~the influence of members of the diffusion network; and (3)~the impact of media outlets when the topics originate in standard news media. Information with high ``novelty'', ``influence'', and ``impact'' can capture huge attention in a short time. Thus, information spread from echo chambers in social media have the capability to create trending topics due to their large scope, like-minded stance, and social influence. Despite social trends containing misinformed statements and false claims, they were presented by social media and reported by mainstream news media~\cite{marwick2017media}. Nowadays, social media companies are making use of this attribute to their profit. They guide the algorithms to select news for its trending topic to keep social media users spending more time on the platform~\cite{carlson2018facebook}. Social media companies, influencers, and news media outlets should take responsibility to carefully display and report social trends. Moreover, social trends can be supervised to detect echo chambers for malicious activities.

\subsection{Political Polarization}
\begin{figure}[hbt!]
\centering
\includegraphics[width=0.5\textwidth]{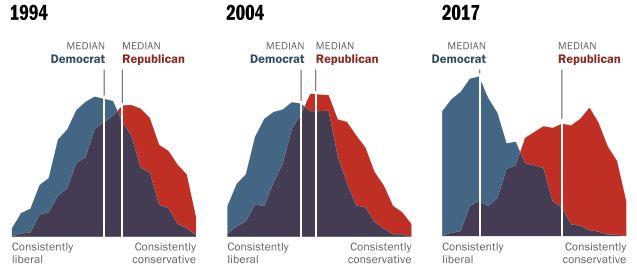}
\caption{Comparison of public political polarization in the U.S over the past two decades, seven Pew Research Centers collected surveys with 10 questions since 1994. Source from Pew Research Center, Washington, D.C. October 20, 2017.}
\label{polarization}
\end{figure}

Political polarization is the divergence of political attitudes to ideological extremes~\cite{dimaggio1996have, fiorina2008political}. According to the evidence from the recent events\footnote{\url{https://www.cfr.org/blog/2020-election-numbers}}, it is clear that the United States experienced record levels of voter engagement. But it also means the country is extremely polarized. Other examples can be found during the COVID-19 pandemic. Based on the US vaccination data from CDC\footnote{\url{https://covid.cdc.gov/covid-data-tracker/\#vaccinations}} as of July 13, 2021, there are 55.6\% of total population who get at least one dose of COVID-19 vaccine. However, a recent study indicate the anti-vaccination movement is currently on the rise~\cite{germani2021anti}. They demonstrate that anti-vaccination supporters are more engaged in discussions on Twitter and share their contents from a pull of strong influencers. As this process evolves by the echo chambers in social media, the community becomes polarized. As shown in Figure~\ref{polarization}, the gap between two major parties in the US has increased while the overlapping has decreased significantly over the past two decades.

In social media, we can observe two giant partisan echo chambers represent two major political groups of people with opposite political opinion and stances~\cite{colleoni2014echo}. Given that individuals tend to align with those who are like-minded in nature, politicians and parties intentionally reinforce the partisan bias inside echo chambers, leading to an increasing level of political polarization~\cite{Levy2019EchoEconomic}. For example, Levy et al.~\cite{Levy2019EchoEconomic} illustrated that politicians made decisions and policies motivated by political purposes rather than social benefits. Political polarization can cause extreme selective exposure, cognitive bias, and correlation neglect~\cite{sears1967selective}. However, Dubois et al.~\cite{Dubois2018Overstated} found that there is only a small segment of the population are likely to find themselves in an echo chamber. Essentially, the impact of partisan echo chambers is overstated. They suggested that echo chamber researchers should test the theory in the realistic context of a multiple media environment.  

\subsection{Emotional Contagion of Users}
Emotional states can be transferred to others via emotional contagion, leading people to suffer from the same emotions without their awareness~\cite{fowler2008dynamic, rosenquist2011social, karami2021profiling}. A recent study showed that extreme emotions are exposed and amplified by echo chambers~\cite{wollebaek2019anger}. This manifestation is usually caused by users who continually receiving misleading contents and conspiracy theories. For example, from a COVID-19 case study of China, Ahmed et al.~\cite{ahmed2020epidemic} illustrated that young people, aged 21-40 years old, were suffering from psychological problems during the COVID-19 epidemic. This is because young people who frequently participate in social media repeatedly receive broadcasts of fatality rate, confirmed cases, and misleading information via echo chambers. Moreover, Del et al. found that inside the echo chamber, active users appear to become highly emotional relative to less active users~\cite{del2016echo}. Their analysis indicated that the higher involvement in the echo chamber enables more negative mental behaviors. Kramer et al.~\cite{kramer2014experimental} provided experimental evidence that emotional contagion can occur without direct interaction between people, and in the complete absence of nonverbal cues. These types of echo chambers are difficult to detect in social media via content-based or network-based methods. 

\section{Echo Chambers Mechanisms}
\label{sec:3_mechanisms}
\begin{figure}[hbt!]
\centering
\includegraphics[width=0.5\textwidth]{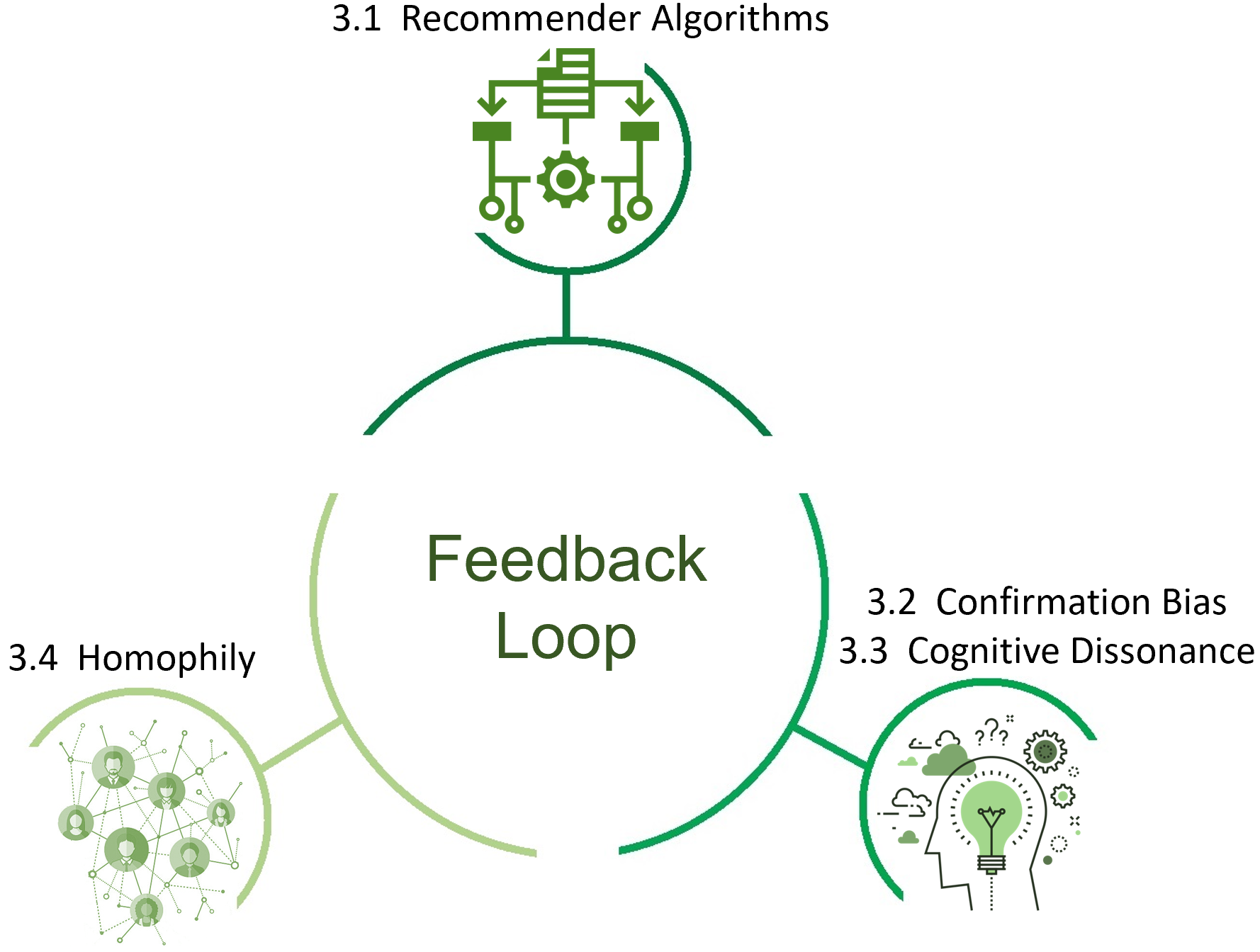}
\caption{Primary mechanisms underlying echo chamber effect related to three main factors: automatic systems such as recommender algorithms (Section~\ref{subsec::RecomAlgo}), human psychology such as confirmation bias and cognitive dissonance (Sections~\ref{subsec::ConfBias} and~\ref{subsec::CogDiss}), and social networks such as homophily (Section~\ref{subsec::Homophily}). These mechanisms are not mutually independent but highly correlated in a way that ultimately create feedback loops that further reinforce the existence of each other.}
\label{mechanism}
\end{figure}

In this section, we discuss the primary mechanisms underlying echo chambers, as shown in Figure~\ref{mechanism}. Specifically, the echo chambers mechanisms consist of three aspects that are connected in a feedback loop: recommender algorithms related to automatic systems, confirmation bias and cognitive dissonance related to human psychology, and homophily related to social networks~\cite{jiang2021mechanisms}. 

\subsection{Recommender Algorithms}\label{subsec::RecomAlgo}
Recommender algorithms trap users into personalized information by using their past behaviors to tailor recommendations to their preferences~\cite{Rastegarpanah2019Fighting}. These prediction engines ``constantly create and refine a theory of who you are and what you will do and want next''~\cite{Pariser2011Filter}, which then forms a unique universe of information around each of us. For example, when clicking on a news article, we show our interest in articles on this topic. The recommender algorithms take note of our behavior and will present more articles about similar topics in the future. As the process evolves, we are getting more and more personalized information, which ultimately leads us to: (1)~becoming the only person in the formed universe, (2)~not knowing how information is recommended, and (3)~unable to choose whether to enter this process~\cite{Pariser2011Filter}. This self-reinforcing pattern of narrow exposure and concentrated user interest caused by recommender algorithms is an important mechanism behind the echo chamber effect. 

Among the many outcomes of such recommender algorithms, e.g., narrower self-interest, overconfidence, decreased motivation to learn, the likely exacerbated polarization has the most negative impact. For this reason, many researchers have criticized recommender algorithms for the increase in societal polarization~\cite{Rastegarpanah2019Fighting,Hannak2013Measuring,Ge2020Understanding}. For example, Dandekar et al.~\cite{Dandekar2013Biased} showed how many traditional recommender algorithms used on internet platforms can lead to polarization of user opinions in society. Therefore, an important line of research studies how to diversify the recommendation results, e.g.,~\cite{jiang2019degenerate}.

\subsection{Confirmation Bias}\label{subsec::ConfBias}

Confirmation bias is the tendency to seek, interpret, favor, and recall information adhering to preexisting opinions~\cite{nickerson1998confirmation}. According to the selective exposure research~\cite{frey1986recent}, we tend to seek supporting information while avoiding challenging information.
Echo chambers are among one of the many outcomes of confirmation bias. The rampant use of social media further amplifies the effect of confirmation bias on echo chambers. There are three types of confirmation bias: biased search for information~\cite{mynatt1978consequences}, biased interpretation of information~\cite{lord1979biased}, and biased memory recall of information~\cite{hastie1986relationship}. In the context of social media, for example, users not only actively seek news that is consistent with their current hypothesis but also interpret information in their own ways. Even if both the collection and interpretation are neutral, they probably remember information selectively to reinforce their expectations, i.e., \textit{selective recall effect}~\cite{hastie1986relationship}.  

Confirmation bias and the provision of recommender algorithms create a self-reinforcing spiral. As described in Figure~\ref{mechanism}, on one hand, recommender algorithms provide users with more of the same content based on their past behaviors to shape the future preference; on the other hand, users accept and even actively seek such information due to confirmation bias. The feedback loop between recommender algorithms and human psychology eventually leads to an echo chamber that shifts users' world view.

\subsection{Cognitive Dissonance}\label{subsec::CogDiss}
In the field of social psychology, cognitive dissonance refers to an internal contradiction between two opinions, beliefs, or items of knowledge~\cite{Festinger1957ADissonance}. For example, if someone eats meat but at the same time cares about the animals' life~\cite{Loughnan2014TheAnimals}. On the grounds that people strive towards consistency, they psychologically feel the pressure to reduce or eliminate the distress caused by dissonance. Festinger~\cite{Festinger1957ADissonance} introduced three major strategies for dissonance reduction: (1)~change one or more of the beliefs, opinions, or behaviors, (2)~increase consonance by acquiring new information or belief, (3)~forget or reduce the importance of the cognitions. Echo chambers are considered as one of the practices in reducing dissonance. People try to seek out for ideologically consonant platforms and interactions to avoid contact with individuals that confront their ideas~\cite{Evans2018OpinionChambers}. Moreover, ideological homogeneity in online echo chambers can encourage extremism. There are two aspects for this stimulation: (1)~one's commitment to their thought will increase dramatically if it has been written down and disseminated to a public audience~\cite{Cialdini2007Influence:Persuasion}. For example, the act of tweeting or posting contents on social media websites; and (2)~discussion with like-minded individuals as well as the social support will reinforce the correctness of that belief~\cite{frey1986recent}. For instance, liking tweets/posts and retweeting/reposting thus boosting attitude extremity~\cite{Bright2020EchoViews}. All of which are in support of decreasing individuals' cognitive dissonance.

\subsection{Homophily}\label{subsec::Homophily}
Homophily, also known as love of the same, is the process by which similar individuals become friends or connected due to their high similarity~\cite{Zafarani2014SocialIntroduction}. This similarity can be of two types: (1)~status homophily, and (2)~value homophily~\cite{McPherson2001BirdsNetworks}. Status homophily deals with people who are connecteddue to similar ascribed (sex, race, or ethnicity) or acquired characteristics (education or religion). Value homophily involves grouping similar people based on their values, attitudes, and beliefs regardless of their differences in status characteristics~\cite{McPherson2001BirdsNetworks}. Depending on the echo chamber's ideology, the echo chamber can be formed due to status homophily, value homophily, or both.
Social media and other online technologies have loosened the basic sources of homophily such as geography and allowed users to bind homophilous relationships on other dimensions like race, ethnicity, sex, gender, and religion.
Moreover, homophily has predictive and analytic power on social media and can be measured by how the assortativity, also known as social similarity, of the network has changed over time and modeled using independent cascade
models (ICM)~\cite{Zafarani2014SocialIntroduction}. By measuring political homophily on Twitter, we can investigate whether the structure of communication is an echo chamber- or public sphere-like~\cite{colleoni2014echo}, or whether there is a homophilous difference between the echo chambers of conservative individuals and liberal ones~\cite{Boutyline2017TheNetworks}. 

As illustrated in Figure~\ref{mechanism}, current recommender systems especially collaborative filtering methods use the concept of homophily to track the effect of user's friends or a crowd of users with the same interest to provide a useful recommendation~\cite{beigi2018similar}. On the other hand, to support their preexisting opinion (i.e. confirmation bias) people tend to follow like-minded individuals and create a homophilous relationship. The feedback loop between these three components will create an echo chamber behavior on social media.

\section{Echo Chambers Detection and Modeling}
\label{sec:4_detection}

Detection of echo chambers is important for several reasons. First, people involved in different echo chambers are ignorant
of information outside. This may lead to the spread of misinformation. Second, detecting different echo chambers may be
instrumental in finding communities with extremist and harmful ideologies. There are many instances where opinions inside 
echo chambers lead to adverse consequences for society as a whole. An example of this is the issue of COVID-19 vaccine hesitancy. 
If individuals inside an echo chamber are unwilling to be vaccinated, they are more likely to contract and suffer sever harm from the virus than those who are vaccinated. Simply put, some beliefs circulating in online echo chambers can have dangerous and lethal real world consequences. 
Cossard et al. look at Italian Twitter to analyze the debate surrounding vaccinations. They conclude that both ardent supporters and critics of these vaccines are at the center of the detected echo chambers~\cite{cossard_falling_2020}.

There are several different strategies employed in computer science and social science literature to detect echo chambers. Since individuals inside an echo
chamber have low intra-community interactions and high inter-community interactions as mentioned in Section~\ref{subsec::Homophily}, community detection algorithms are
widely used. Our literature review suggests that the main approaches to detect echo chambers include content based
detection and network based detection. Finally, since ground truth data about echo chambers is scarce, researchers try
to model echo chambers using various techniques based on assumptions about the ground truth.


\subsection{Content-based Detection}

Here we look into methods that utilize the content produced by users to detect echo chambers. Users interact with social media platforms in many ways: posting, liking posts, reblogging posts, or commenting on them. These interactions provide a valuable information about the beliefs and opinions of users that can be used to detect echo chambers.

\begin{figure}[hbt!]
    \centering
    \includegraphics[width=0.5\textwidth]{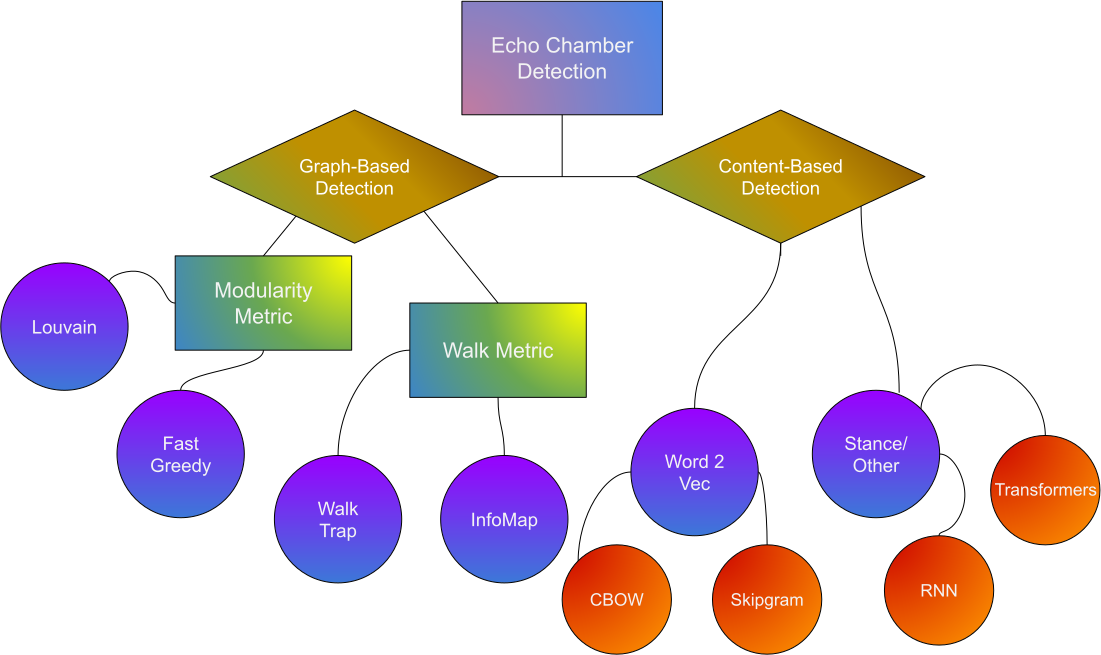}
    \caption{Hierarchy of methods generally used for detecting echo chambers. Modularity and Walk heuristics are used
    to analyze the network while content based methods analyze the text.}
    \label{fig:detection}
\end{figure}

\bigskip
\noindent
\textbf{Stance and Opinion Detection}\
We can use social media users'
opinions (e.g., the tweets they write) to measure the similarity (or dissimilarity) between users. There are many ways
to mine users' opinions, such as simple textual similarity measures like TF-IDF~\cite{ramos2003using},
Word mover~\cite{kusner2015word}, Doc2vec~\cite{le2014distributed}, or stance detection. N-gram analysis and TF-IDF are used to generate features by Nguyen et al.~\cite{nguyen_mitigating_2017} in conjunction with k-means clustering to find echo chambers. The authors propose that using this approach to find echo chambers and cross linking across echo chambers might help slow their growth. 

Stance detection is widely used in fake news and rumor detection work~\cite{Kai2017Fake,kumar_tree_2019}. Traditionally, this task involved finding word-based similarities. With the popularity of neural networks, embedding-based similarities are seeing more usage. In word-based embeddings, the embedding of the narrative in question is
determined by aggregating the embeddings of individual words. The embedding function can be an aggregate function like mean, maximum, or average, etc. On the other hand, neural networks are capable of embedding the entire narrative~\cite{Orbach2020Out}. Once
the embeddings are created, they can be compared together to identify whether they agree or disagree with each other. Measuring user sentiment across communities with differing views might also help us gain insights into echo chambers. Del Vicario et al. use a sentiment analysis API to measure the emotional distance between communities~\cite{del_vicario_mapping_2017}. Given the topic, the greater the community distance, the more polarizing it is. 

Word-based embeddings that are commonly used can either be taken from a source where they are pre-generated, or they can
be generated by the a training corpus. Some of the commonly used pre-generated embeddings are Stanford's GloVe~\cite{pennington2014glove} and Google Word2Vec~\cite{mikolov2013efficient} embeddings. In Word2Vec, the embedding of a
word is created by either a Continuous Bag of Words~(CBOW) or Skip-gram neural network model. These models extract the
embeddings of words by either predicting the word from the surrounding words~(CBOW) or predicting surrounding words
(Skip-Gram).

For embedding an entire narrative Recurrent Neural Networks(RNN) and Transformer-based models are being used extensively
more recently. Apart from vanilla RNNs, which have recurrent
cells that share hyperparameters, there are several other types of RNNs. The most common among them are LSTMs~\cite{Hochreiter1997}
and GRUs~\cite{DBLP:journals/corr/ChoMGBSB14}. For transformer-based models, BERT~\cite{devlin2018bert} and its variations
are used more prominently in echo chamber research~\cite{han2019fallacy}. Transformer models have two components at a high level: an encoder
and a decoder. BERT uses the encoder part of the transformer model only. It is commonly used for generating the encodings
of an input narrative.

\bigskip
\noindent 
\textbf{Opinion Distance}\
The goal of measuring opinion distance is to find if two opinions are similar or not. We input text opinions from
two different users, and the output of the algorithm will be a value between 0 and 1. A distance of 0 indicates that the
two opinions are identical regarding the topics discussed in the two opinions, and 1 represents the opposite.

Compared to the other opinion mining methods, this method can capture the nuances between opinions. In other words,
this method could differentiate between two opinions that have the same stance on a topic (let's say Brexit, for
example); however, they differ in the justification of their opinion (in that case, one believes that Brexit is good for the economy and the other thinks it's terrible but necessary to reduce immigration for example).

Gurukar et al.~\cite{Gurukar2020Towards} developed a method to compute the distance between opinions. This method consists of three steps: (1)~opinion representation, (2)~mapping the opinion subjects between the two opinions, and (3)~computing the
opinion distance. The goal of opinion representation is to identify the opinion topics the users talked about. This step is dynamic, meaning that the user's opinion topics are only based on their opinion text, not on a preset target topic. Therefore, we need a way to map the different topics between opinions. After the mapping step, we compute the
distance between opinions using this formula~\cite{Gurukar2020Towards}:

\begin{equation}
    OD(O1, O2) = \frac{\sum_{(i,j) \in M} f( pol(S_i^1) , pol(S_j^2) ) }{ 2 \times \left | M \right | }
\end{equation}

Here, $f$ is a distance function, $S_i^1$ and $S_j^2$ represent the subjects $i$ and $j$ in opinion $O_1$ and $O_2$ respectively, $pol$ is a function that measures the polarity of the subject towards an opinion. Finally, $M$ is the set of mapped opinions.

\bigskip
\noindent
\textbf{Polarization Detection}\
To detect echo chambers, one could utilize methods used to detect polarization and polarized speech. There are generally two opposing sides for most public discourse topics, one that supports and the other opposes it.
Usually, the two sides use different languages to describe the same idea~\cite{Garimella2017Longterm}. For example, a person might use the term ``global warming'' and someones else uses ``Climategate''. Both opinions deal with the same subject; however, the language they use and their attitude differ.

To analyze political polarization in Twitter, Garimella and Weber~\cite{Garimella2017Longterm} looked for signs of polarization
in the way users interact with each other and their content. Specifically, they studied the ``follow'' patterns, retweet
behavior, and how partisan the hashtags used in tweets are. The same features could be used to detect echo chambers.

The leanings of individuals can help with the detection of echo chambers~\cite{Cinelli2021Effect}. Let the set of content produced
by a user, $i$, be represented by the set $C   _i = \{c_1, c_2, ..., c_{a_i}\}$, where $a_i$ represents the number of activities
by the user. Cinelli et al.~\cite{Cinelli2021Effect} define the leaning of the user as:
\begin{equation}
    \label{eq:simple_leaning}
    x_i = \frac{\sum_{j=1}^{a_i} c_j}{a_i}
\end{equation}
If the distribution of the leanings, $P(x)$ is bi-modal, then it is an indicator of binary polarization. Bessi et al. study the formation of echo chambers by using YouTube links shared on Facebook as a petri dish~\cite{bessi_users_2016}. The likes, shares and comments are used as features to measure how polarized the users are. The authors argue that the bimodal PDF based on the users polarity is evidence for the formation of echo chambers. 

Another interesting niche approach for echo chamber detection is the work by Bessi et al. The authors look at echo chambers from the perspective of Big Five personality traits. An unsupervised approach is used to infer the personality traits from text. The authors conclude that a 
certain combination of the big five personality traits are responsible for individuals ending up in an echo chamber~\cite{bessi_personality_2016}.


\subsection{Network-based Detection}
Looking at the social network can give us insights into echo chambers. Because of homophily, people inside an echo chamber
are well connected while people outside of echo chambers are weakly connected. Apart from connectedness, one of the common
ways to detect echo chambers in fake news publications is to look at the propagation patterns of topics being discussed.

Network based detection is one of the most common ways to locate echo chambers. Because of the increasing use of 
social media platforms, they are becoming habitats for echo chambers. An example of this is demonstrated by Guarino et al. where they use community detection to illustrate that high segregation and clustering of communities by political 
alignment is evidence for the presence of echo chambers in social media~\cite{cherifi_beyond_2020}.

\bigskip
\noindent
\textbf{Fast Greedy}\
The fast greedy algorithm performs hierarchical agglomerative clustering to get a network that optimizes on modularity~\cite{newman_fast_2004}.
Modularity is a measure of the quality of a graph containing communities.
Let $A_{ij}$ be the weight of the edge between $i$ and $j$. Let $k_x$ be the total weight of the edges attached to node $x$.
Let $c_x$ be the community for node $x$. $\sigma(x, y)$ is an indicator function which takes the value 1, if $x = y$ and
0, if $x \neq y$. $m$ is defined as the total number of edges. Modularity, $Q$, is defined as:
\begin{equation}\label{eq:modularity}
Q = \frac{1}{2m} \sum_{i, j} \bigg[ A_{i, j} - \frac{k_i k_j}{2m} \bigg] \sigma(c_i, c_j)
\end{equation}
Fast greedy is one of the simplest modularity based algorithms that can be used for detecting communities. Finding the 
optimal graph configuration for modularity is an NP-Hard problem. Thus, heuristics like the greedy approach are used 
in literature. Fast Greedy is one of many community detection approaches used by Cossard et al. to find echo chambers for vaccine supporters and skeptics~\cite{cossard_falling_2020}.

\bigskip
\noindent
\textbf{Louvain}\
The Louvain algorithm is used to detect communities in a graph~\cite{blondel_fast_2008}. The objective function of this algorithm is modularity (Eq.~\ref{eq:modularity}). The algorithm starts by assigning a unique community to each node. This algorithm iteratively combines communities to gain modularity until convergence. First, each node is compared with the neighboring nodes. Communities are changed to match
the neighbors if it increases the modularity. Each community is reduced to a single node where the new edge weight is a sum of the previous graph.

The popularity of louvain algorithm's usage is testament to its effectiveness. Guarino et al. use Louvain for their 
DisInfoNet toolbox~\cite{cherifi_beyond_2020}.  Cossard et al. use Louvain and other community detection algorithms to
detect echo chambers in an Italian subset of Twitter~\cite{cossard_falling_2020}. Nourbaksh et al. use Louvain on a co-linking network to detect echo chambers~\cite{nourbakhsh_mapping_nodate}.

\bigskip
\noindent
\textbf{WalkTrap} The WalkTrap algorithm is based on the idea that a random walk tends to end in the same community~\cite{pons_computing_2005}.
Let $P_{xy}^t$ be the probability of walking from $x$ to $y$ in $t$ timesteps, $d(x)$ be the degree of $x$. Then the distance, $r_{ij}$ between $i$ and $j$ is defined as:
\begin{equation}
    r_{ij} = \sqrt{\sum_{k=1}^n \frac{(P_{ik}^t - P_{jk}^t)^2}{d(k)}}
\end{equation}

Similar to the Fast Greedy algorithm, the algorithm is initialized by assigning a unique community to each node. Hierarchical
clustering is used to group similar nodes. The Closest communities are merged and form a new graph where each node is a merged community. The process is repeated until we have one community. In order to deal with the high computational complexity of finding the optimal communities, a Monte Carlo approach can be used to estimate the probabilities for the random walks. 

An example of WalkTrap algorithm for detecting echo chambers is illustrated by Del Vicario et al.. The authors of this 
work wanted to study the opinions on Brexit by looking at Facebook data. They created a Bipartite graph and used community detection to find echo chambers within the data~\cite{del_vicario_mapping_2017}.

\bigskip
\noindent
\textbf{Infomap}\
The infomap algorithm optimizes the map equation~\cite{rosvall_map_2009} to find communities. The map equation tries to
find the lower bound on the length of the sequence used to represent a walk on the graph. The representation of the walk can be minimized by using Huffman codes. The most frequently visited nodes may be represented by a lower number of bits. In order to minimize the length of the walk, the graph can be divided into different modules. Each module has a codebook (module codebook). There is also a codebook representing movement between the modules (index codebook). The
description length for a module may be represented by the map equation:

\begin{equation}
    L(M) = q_\curvearrowright H(\mathcal{Q}) + \sum_{i = 1}^{m}{p_{\circlearrowright}^i H(\mathcal{P}^i)}
\end{equation}

Communities detected by the map equation may differ from those formed by a modularity based approach. This is because the idea behind the map equation is to optimize the flow of information while the modularity is a connection-based metric.

InfoMap is another algorithm that is popular among the research community for studying echo chambers. It is one of
the algorithms used by Cossard et al. for studying echo chambers related to vaccination skepticism~\cite{cossard_falling_2020}. It is also the only community detection algorithm used by Du et al. in their 
work where they study how echo chambers strengthen by comparing detected communities across two time points~\cite{du_echo_2016}. 

\bigskip
\noindent
\textbf{Other Approaches}\
There are a plethora of other approaches for detecting communities. Spectral approaches, for instance, use eigen decomposition of the adjacency matrix to perform clustering. Heat kernel based detection~\cite{kloster_heat_2016} is inspired from heat diffusion. Shi et al.~\cite{shi_comparison_2020} compare 70 community detection approaches on basis of their quality and runtime. Brugnoli et al.~\cite{brugnoli_recursive_2019} use time series algorithms to study 
subclusters within echo chambers.

\begin{figure}[hbt!]

\centering

\begin{subfigure}[b]{0.5\textwidth}
  \includegraphics[width=1\linewidth]{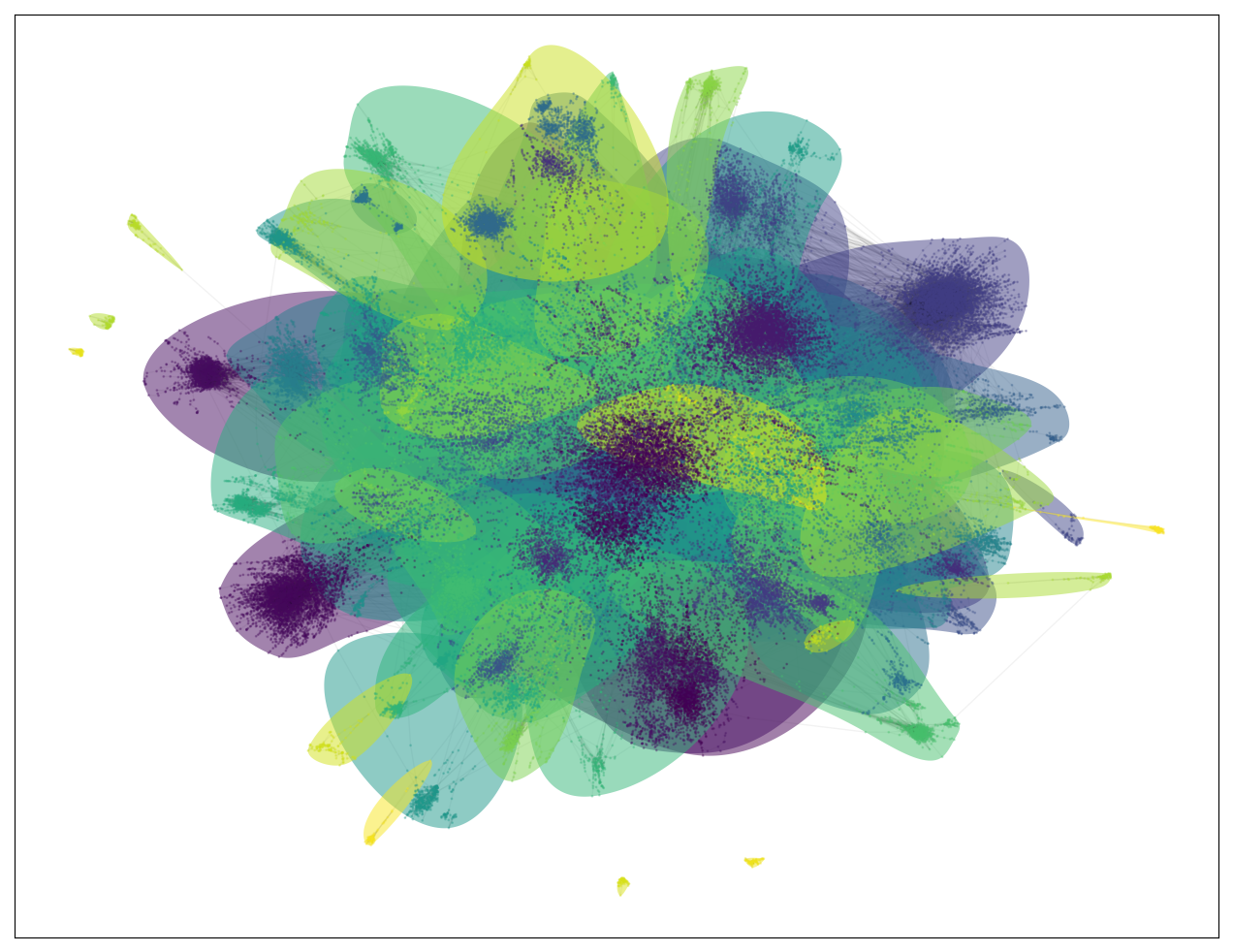}
  \caption{Louvain Communities on Twitter}
\end{subfigure}

\begin{subfigure}[b]{0.5\textwidth}
  \includegraphics[width=1\linewidth]{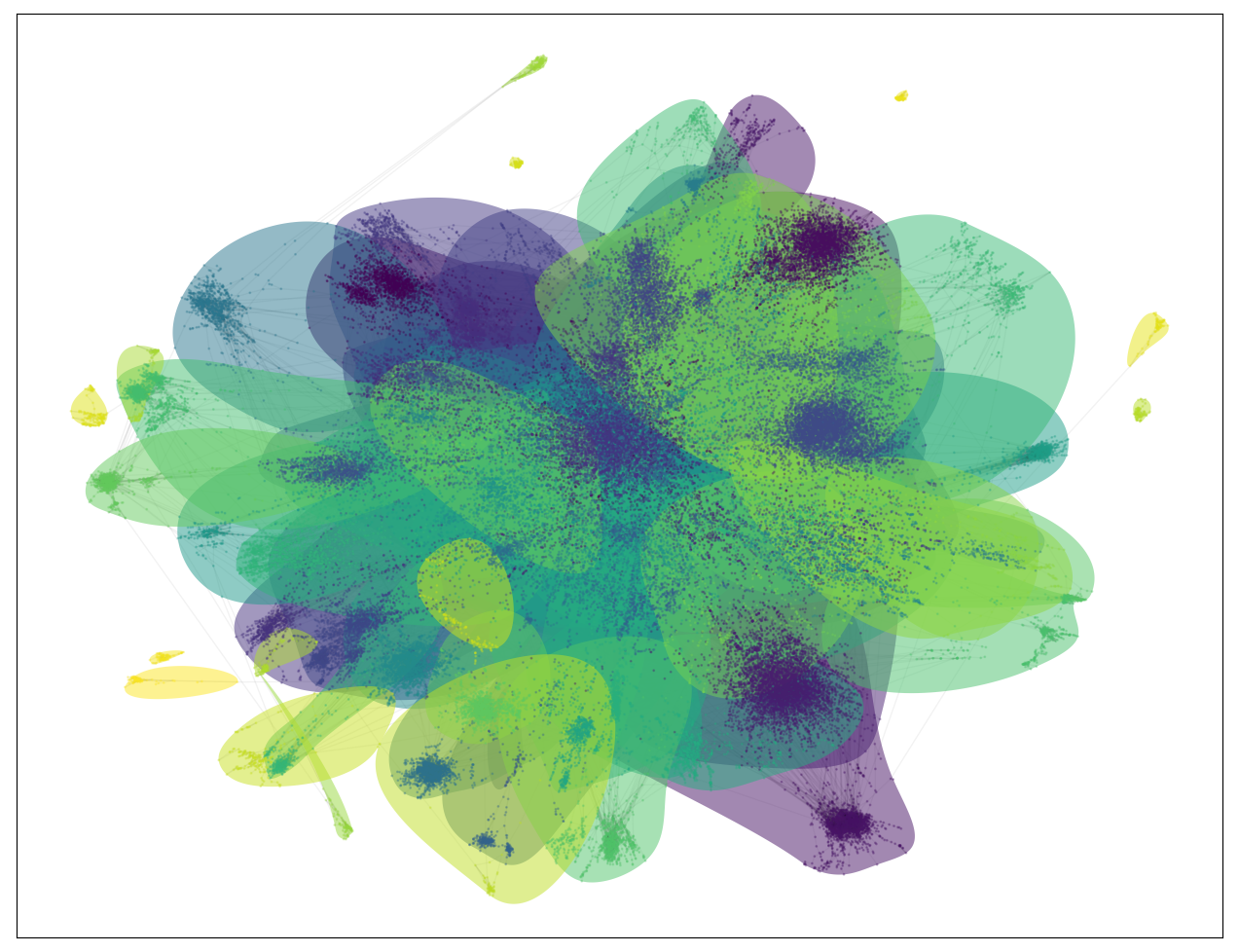}
  \caption{Infomap Communities on Twitter}
\end{subfigure}

\caption{Comparison between the communities detected by Louvain(Modularity-based) and Infomap(Walk-based) algorithms over a twitter network of ~3M tweets. Over large networks community detection algorithms are victims to the heterogeneity of interactions.}
\label{fig:community-detection}

\end{figure}

\begin{table*}[t]
\begin{tabular}{ 
  | >{\centering\arraybackslash}p{0.1\textwidth} 
  | >{\centering\arraybackslash}p{0.17\textwidth} 
  | >{\centering\arraybackslash}p{0.18\textwidth} 
  | >{\arraybackslash}p{0.43\textwidth}   | }
\hline
\textbf{Algorithm}         & \textbf{Papers} & \textbf{Metric} & \textbf{Description}      \\ \hline\hline
Fast Greedy       & \cite{cossard_falling_2020} \cite{del_vicario_mapping_2017} & Modularity &  Combine communities greedily to maximize modularity\\
\hline
Louvain           & \cite{cherifi_beyond_2020} \cite{cossard_falling_2020}  \cite{Cota2019Quantifying}  \cite{nourbakhsh_mapping_nodate} & Modularity  & Modularity maximization using optimization and aggregation phases                    \\
\hline
InfoMap           &  \cite{du_echo_2016} \cite{cossard_falling_2020}     & Walk (code length)    & Shortest representation of Walk paths \\
\hline
WalkTrap          &  \cite{del_vicario_mapping_2017}   & Random Walk & Likelihood of Random walk ending in the same community\\
\hline
\end{tabular}
\caption{Common Graph-based community detection algorithms and their usage in literature for detecting echo chambers.}
\end{table*}



\subsection{Echo Chamber Modelling}\

Due to the difficulty of studying echo chambers based on real-life data, some researchers use a different method to model echo chambers and study the information propagated in them. In table 2, list some of these methods, and we explain how they work.

\bigskip
\noindent
\textbf{Friedkin-Johnson Dynamics.}\
The Friedkin-Johnson Dynamics~(FJ) model can be used to simulate the leaning of different users in a social network~\cite{Chitra2020Analyzing}.
This model assumes that the leanings of each user on each issue can be represented as a spectrum from -1 to 1, where -1
represents one extreme and 1 represents the opposite extreme e.g. agreement or disagreement on a particular issue. Each
user is represented by a node in a graph, $\mathcal{G}$. The weight of the edges, $w_{ij}$, in the graph signify the
influence they have over each other. Since the model is represented by an undirected graph, the influence is reciprocal.
The crucial difference between the FJ model and leanings as defined in Eq.~\ref{eq:simple_leaning} is that this model
assumes that there are certain innate opinions that the users have that cannot be changed. However, as time passes each user builds on these innate opinions based on the influence of peers.

The FJ model uses a time series where the leaning, $z_i^t$, of a user, $i$, in the social network at time $t$ is a function of
their leaning,$z_i^{t-1}$ , at time step $t-1$ and their innate leaning $s_i \in [-1, 1]$. Let $d_i$ be the degree of
user $i$. The leaning at time $t$, is formally defined as:
\begin{equation}
    z_i^t = \frac{s_i + \sum_{j \neq i} w_{ij} z_j^{t-1}}{d_i + 1}
\end{equation}

One of the advantages of using the FJ model is that it is simple to find where the leanings of the users will end up as
time progresses. The state at which the leaning does not change~(equilibrium) is simply denoted as:
\begin{equation}
    z^* = \lim_{t \rightarrow \infty} z^t
\end{equation}

The model is capable of representing concepts like polarization, local/global disagreement, and local/global conflict~\cite{Chitra2020Analyzing}.

\bigskip
\noindent
\textbf{Bounded Confidence Model.}\
This Model is used to predict the opinions of agents by modelling their interactions~\cite{deffuant_mixing_2000}. Instead of having binary opinions on an issue, each entity can have an opinion in a spectrum
on $[0,1]$. The change in opinions is determined by the threshold, $d$, and the convergence parameter, $\mu$. The idea here is that if two entities with different opinions interact, they are unlikely to change opinions if
they differ by the threshold, $d$. If their opinions are below the threshold, they adjust their opinions based on the
barycentric combination determined by the factor $\mu$. Thus, the model concentrates the opinions of entities towards
centroids. The number of centroids might reflect the different echo-chambers that have been formed.

\bigskip
\noindent
\textbf{Stochastic Block Model.}\
The Stochastic Block Model~(SBM) creates a graph model of a community where the parameters of the model define how much members of a community intermingle inside and outside their community~\cite{abbe2017community}. This model is useful when combined with the assumption of homophily, i.e. people with similar leanings tend to be connected. It must be noted that this model does not simulate homophily by default.

SBM creates a random graph based on a number of parameters. The basic assumption is that there are two or more
communities created from $n$ nodes. For each pair of nodes, let $p$ be the probability that they are connected given that they
are in the same community. Let $q$ be the probability that they are connected if they are in different communities. This
results in a graph over both the communities. In order to simulate homophily, we set $p$ to be greater than $q$. This
will result in nodes of the same leanings/communities being connected more inside the communities.

SBM can be useful when combined with FJ model to study the effects of Filter Bubbles~\cite{Chitra2020Analyzing}. FJ model is useful
to determine the equilibrium leanings of individuals while SBM can provide us with a way to model the initial state of
the social network. We can simulate filter bubbles by adding or removing edges.

\begin{table*}[t]
    \centering
    \begin{tabular}{lll}
        \toprule
        & \textbf{Propogation Criteria} & \textbf{Representation} \\ 
        \cmidrule(l){2-3}
        \textbf{FJ Dynamics}        & Innate/Evolving Leaning       & Graph             \\
        \textbf{Bounded Confidence} & Evolving Leaning              & Grid              \\
        \textbf{Stochastic Block}   & Intermingling                 & Graph/Communities \\
        \textbf{Cascade}            & Discourse Interactions        & Tree              \\
        \textbf{Epidemic}           & Adjacency                     & Grid              \\ 
        \bottomrule
    \end{tabular}
    \caption{Models typically used for Echo Chambers and Filter Bubbles}
    \label{tbl:models}
\end{table*}

\bigskip
\noindent
\textbf{Cascade Models}\
Graph models can be used to study echo chambers in social media. Since social media discourse is temporal, the propagation
patterns can be encapsulated in a cascade model~\cite{zhou_fake_2018}. The cascade model is represented using a tree. Each
node in the tree is a user in the social network. The root node represents the user who began the discourse. Each level
of the tree represents propagation. This could be a share, reply, or any other kind of social media-specific interaction
in which information is passed from one user to another. There are multiple variations of a cascade propagation model.
Cascade similarity between several disconnected discourses might help with the detection of echo chambers.

\bigskip
\noindent
\textbf{Epidemic Models.}\
Epidemic models are inspired by the spread of disease. These models are used in publications to detect echo chambers~\cite{Cinelli2021Effect}.
The model represents individuals as nodes with one three states: Susceptible(S), Infected(I), or Recovered(R). Each individual has a certain probability of moving from one state to another based on the state of their neighbors. There are different variations of this model based on the possible transition~(such as SIR, SIS, SIRD, and SEIR). The latter two variants add additional states to the model like Deceased(D) and Exposed(E).

\section{Prevention and Mitigation}
\label{sec:5_prevention}
In this section, we discuss how to prevent echo chambers on social media from forming. We further discuss how to mitigate echo chambers in cases where they are already formed. We divide prevention and mitigation strategies into two types: algorithm-focused and human-focused. The algorithm-focused strategies try to address the causes of echo chambers that occur due to algorithmic curation and content recommendation. The human-focused strategies include methods designed to give the user more power over their information environment by urging them to think about the quality of the information they consume. 

\subsection{Algorithmic Prevention}

In recent years, there is a focus on the fairness of machine learning algorithms as a response to the realization that some of these algorithms might lead to the unfair representation of some users based on some protected attributes such as gender, race, or age.  We argue that echo chambers' problem could be thought of as a fairness problem if we redefine the problem as it is not fair to users to be recommended only content or other users that excluded diverse opinions. As we mentioned in previous sections, the source of many problems related to echo chambers is recommender systems. They could produce a polarized network (echo chamber) of users by recommending homophilic links, or produce the filter bubble effect by suggesting biased content. To solve this problem, we examine research field of fairness in recommender systems. There are three main types of solutions to the problem of unfair recommender systems: preprocessing solution,  in-processing, and post-processing. 

We start with preprocessing solutions where we try to address the polarization in the data used to train recommendation algorithms. Inspection of data poisoning attacks from Rastegarpanah et al.~\cite{Rastegarpanah2019Fighting} suggest adding additional training data. They refer to these data as antidote data. This method has the advantage of not modifying the training data nor modifying the recommendation algorithm. A small amount of antidote data (1\% more new data samples) reduces polarization of the recommendations by 50\%. However, decreasing polarization will lead to a decrease in recommendation accuracy, which is an expected result since the ``correct'' recommendation created the polarization in the first place.  

In-process methods try to modify the way the recommendations systems works to prevent the formation of echo chambers. One way to prevent creating echo chambers is by modifying the objective function of the recommender system. Chitra and Musco~\cite{Chitra2020Analyzing} showed that a simple modification to the objectivate function mitigates the filter bubble effect. By adding a regularization term to the objective function, Chitra and Musco showed that the recommendation algorithm would take into account the whole network instead of focusing on the user. 

Post-processing methods can be thought of as the methods that we use after the recommendation has been made with or without any changes to the data or the algorithm.  One suggested idea is to detect a counter-argument to each polarized argument. Orbach et al.~\cite{Orbach2020Out} try to identify, from among a set of speeches on the same topic and with an opposing stance, the ones that directly counter it.

Another important body of work that merits mention in the discussion of algorithmic prevention of echo chambers is that done by the Polarization lab at Duke University. The lab conducted a set of studies on political polarization and echo chambers, and summarized them in the book \emph{Breaking the Social Media Prism}~\cite{Bail2021}. The results of the studies concluded that exposing users to polar opposite beliefs in order to draw them out of their echo chambers actually further entrenched them in their beliefs. The lab did find an effective approach in depolarizing users by exposing them to viewpoints only slightly less polar than the ones they held. Essentially, going one bandwidth lower in acceptability for beliefs.

These findings have deep implications for algorithmic prevention of echo chambers, should the echo chamber already exist. Algorithms seeking to draw a user out of their echo chamber by depolarizing the user must not only analyze and decide what content to recommend, but must also decide whether the content is in the latitude of acceptability that would slightly depolarize the user. This is certainly a difficult challenge, but the results could bear fruit for specifically the mitigation of echo chambers.

\subsection{Human-focused prevention} 
A secondary method of echo chamber prevention and mitigation relies on giving users more power to curate their own information feed with an eye towards less bias. We refer to methods that take this route as human-focused prevention methods. These types of methods have been tried publicly by Facebook and Twitter, to varying degrees of success. The main human-focused methods are currently: labeling of misinformation/Fake News, fact checking, and ``nudging'' the user to think about accuracy. 

All of these methods, while well intentioned, face certain limitations. labeling bad information can potentially lead the a member of an echo chamber to realize that they are caught in a chamber. Conversely, labeling could result in the \emph{implied truth effect}~\cite{impliedtruth}. The implied truth effect means that attaching labels to fake news or misinformation increases perceived accuracy of headlines without labels. So, if a user in an echo chamber were to be subject to the implied truth effect, it would mean that any piece of bad information that slipped labeling would be perceived as more accurate by the user. Fact-checking also runs into resistance from the underlying features of echo chambers, primarily the distrust of help from outside the echo chamber. Indeed, fact-checking could easily backfire and entrench the user more in their beliefs. Mosleh et al. found that fact-checking and debunking led users to abuse the fact-checker with toxic comments~\cite{mosleh_toxic}. Fact-checking faces further challenges because it relies primarily on human agents to check a piece of information. Fake news and misinformation spread much faster than true information ~\cite{Vosoughi2018}, and in an echo chamber, fact checkers may not be able to keep up with the velocity of information spread. While some studies have found success with fact-checking~\cite{Gillani2018Me}, the results are a mixed bag and require further research before fact-checking can be considered an effective human-focused approach to prevention and mitigation of echo chambers. 

Utilizing the ``nudge'' approach has displayed some promising early results. The basis of the nudge approach comes from the concept of ``nudging'' an agent towards a desirable outcome. Pennycook et al. found that susceptibility to fake news can be explained in part by a simple lack of reasoning on the part of the user~\cite{pennycook2019lazy}. With respect to echo chambers, nudging the user to think about accuracy could potentially reduce the negative effects of the chamber. Two separate studies have found that nudging users on social media to think about accuracy of information can yield positive results in reducing the belief in fake news and misinformation~\cite{deliberationreduces,pennycook2021shifting}. These results could be of use in echo chamber prevention and mitigation because they counteract two features of echo chambers: the resistance of information that disagrees with the echo chamber member's worldview, and the awareness that the member is in an echo chamber at all. 

Clearly, a human-focused approach to echo chamber prevention and mitigation is a difficult and multifaceted problem. The mixed results from current approaches makes it abundantly clear that there are many opportunities for research in this area. Future work can build on the results we have discussed here, and take advantage of the dearth of psychological studies to develop an approach that ultimately proves beneficial to the echo chamber member.

\section{Challenges and Open Problems}
\label{sec:6_challenges}
Though complicated, echo chambers are a problem worth researching because of their prevalence on social media and their wide adoption across multiple platforms. In our estimation, the source of challenges stemming from echo chambers is the fact that echo chambers have many shareholders: (1)~the members of the echo chamber, (2)~the social media platforms, and (3)~the ``offline'' world. Each one of these shareholders introduces challenges as well as open problems to solve. This section discusses these aforementioned shareholders' challenges and other challenges and the open problems related to echo chambers.

\subsection{Human-Related Challenges}
The human element of the echo chamber makes solving and studying it a challenging problem. Any work related to echo chambers and polarization should consider how people inside an echo chamber consume content, how they perceive the world, and how they view people outside their echo chamber. Echo chamber members have four critical features that contribute to this problem: (1)~they are not aware of their echo chamber, (2)~they select contents that adhere to their beliefs, (3)~they resist any information that disagrees with their worldview, and (4)~they distrust help from outside the echo chamber. Following, we are going to go over each one of these features briefly.

Echo chamber members are generally not aware of their echo chamber. Most users would not suspect that the content they consume and their relationships online are part of an echo chamber. Raising awareness about echo chambers and their effect (See Section~\ref{sec:2_attributes}) on the individual and society is an essential step towards solving the problem of echo chambers. The presentation of the information that leads users out of their echo chambers and towards a more civil online presence is critical. The presentation should be done carefully and in a way that avoids belittling users and their beliefs, to avoid the toxic abuse mentioned in Section~\ref{sec:5_prevention}. A possible way to combat echo chambers is by designing a tool that helps users find whether they are in an echo chamber. One example of such a tool is the Check-my-echo\footnote{\url{https://www.polarizationlab.com/echo-checker}} tool by the polarization lab at Duke University. 

As we mentioned in Section~\ref{sec:2_attributes}, confirmation bias and selective exposure are core factors for forming echo chambers. Users will select the information that reinforces their existing beliefs due to confirmation bias. Also, they will avoid information that contradicts their beliefs. Bakshy et al.~\cite{Bakshy2015Exposure} found that individuals' choices play a more decisive role in limiting exposure to cross-cutting content that conflicts with the echo chamber's worldview.  Therefore, any method that tries to recommend new content outside the echo chamber should consider this fact and not recommend content that could cause echo chamber members to double down on their preexisting beliefs. 

Echo chambers members are more involved in producing~\cite{Dube2015Vaccine} and consuming~\cite{Schmidt2018Polarization} content related to their ideology than any other groups outside the echo chamber. This fact leads us to conclude that echo chamber members' attitudes are rooted more deeply in a person's social and psychological background~\cite{Schmidt2018Polarization} than people outside the echo chamber. Therefore, their stance on different topics is firm and less likely to be changed in a short period of time. We must realize that the echo chamber problem is due to the interaction of the structure of social media with select elements of human psychology. Solving it will take a fundamental change in the way social media works. 

Members of echo chambers tend to distrust any voices outside their group that offer information that conflicts with their worldview. In addition this distrust, being confronted with opposing information can reinforce preexisting beliefs, and even well intentioned debate can lead to an increase of opinion polarization~\cite{Schmidt2018Polarization}. Trying to bridge the echo chamber by sharing diverse content results in a loss in network centrality and content appreciation of users that attempt to solve the problem~\cite{Garimella2018Political}. Therefore, the type of corrective information and how it gets presented is a critical concern for solving the echo chamber problem.  
\subsection{Social Media-Related Challenges}

The main objective of social media platforms is to keep the user engaged with the platform by spending more time consuming content, and therefore consuming more advertisements. This is the fundamental principle driving revenue and profit for mainstream social media platforms. However, we believe that social media platforms must balance between the social good and their financial profit, which is a legitimate concern. Although recently social media platforms have introduced some measures to combat the filter bubble and fake news problems, these methods have found limited success. We posit that the main way for platforms to reduce the echo chamber problem is by designing an echo chamber-aware recommender system. As we mentioned in Section~\ref{sec:3_mechanisms}, recommender systems are one of the leading perpetrators of creating echo chambers. Recommender systems are utilizing users' own biases to personalize recommendations in order to keep them engaged with social media. Therefore, the ethical responsibility lies on recommender systems' designers to consider the effects of their systems on the formation of echo chambers. In accord, any solution to echo chambers must ensure the quality of the content (or connections) that is recommended to users. What we mean by the quality of the recommendation is that any content that is recommended to users must seek to avoid the formation of echo chambers and opinion polarization and at the same time ensure that social media platforms meet their obligations to keep users engaged on the platform. To design echo chamber-aware recommender systems, we need to find the right balance between the quality of recommendations and the main objective of such systems, which is to recommend content that interests users. These recommender systems should seek to be aware of the overall polarization of the whole social network and predict how their recommendation affects the network's overall polarization. One possible way to design such systems is to make recommendations that make sure that users are informed about other viewpoints other than theirs. This method is inspired by AllSides\footnote{\url{https://www.allsides.com/unbiased-balanced-news}}, a website that shows users a news article and a brief extract of news articles from three different political views right, center, and left.

\subsection{Content Moderation-Related Challenges}
With the dearth of fake news and misinformation and the increase in political polarization, many have called on social media companies to have better oversight over their platforms. Consequently, many social media platforms have started to introduce measures to combat fake news and abusive content. For example, Twitter appended a fact-checking notice on President Trump's tweets that Twitter judged as misleading or false claims. This type of measure is called \textit{content moderation}. Although these efforts are important ways to combat echo chambers and misinformation, their effects are not well understood. More work is needed to develop methods and suggestions of best practices to moderate content in social media platforms.  We identified these problems as potential concerns: (1)~the sheer amount of content that gets posted on social media make unbias, and objective human moderations near impossible, (2)~automatic content moderation do not work as good as needed to stop abusive and polarizing content, (3)~the subjective and potentially bias human moderation could lead to the formation of a network-wide echo chamber, and (4)~excessive moderation could make some users feel unwelcome, leading them to leave the social media platform, which causes financial losses and might accelerate the formation of a network-wide echo chamber. 

We call social media that has only one ideological leaning a \textit{network-wide echo chamber}. For example, Reddit is a left-leaning echo chamber, while Gab is a right-leaning echo chamber~\cite{Cinelli2021Effect}. Although we do not believe that content moderation (especially when it comes to Reddit) is the sole cause of this phenomenon, content moderation contributes to the feeling that some ideological or political leanings are not welcome in ``mainstream'' social media. For example, in the United States, some conservatives believe that the ``mainstream'' social media has a bias against them. This belief causes the emergence of social media that has a conservative-leaning, e.g., Gab, Parler, and Rumble, to name a few. Having network-wide echo chambers is a very concerning indicator of how politically polarized society is right now. 


\subsection{Political Environment-Related Challenges}
Society's increasingly polarized environment affects social media, which in turn contributes to the polarization of society, creating a feedback loop of polarization~\cite{Jasny2015Empirical}. In this vicious cycle, echo chambers create and spread polarized content. The polarized content increases the political polarization of society and political discourse. This polarized environment helps echo chambers to grow and accelerate the formation of new echo chambers. Breaking this vicious cycle is challenging, and the social and political climates are very susceptible to this type of interaction. DellaVigna and Kaplan~\cite{DellaVigna2007Fox} concluded that Fox News affected voter turnout and the Republican vote share in the Senate.  Moreover, De-Wit et al.~\cite{DeWit2019social} mention that the political base affects politicians' tweets, which reporters use to build their news headlines.

Rancorous political discourse is contributing to the echo chamber problem and leads to an increase in opinion polarization because echo chamber members are more involved with their content, and they show signs that their beliefs are more deeply rooted~\cite{Schmidt2018Polarization}. In fact,  two-sided neutral arguments have weaker effects on reinforcement than one-sided confirming and contradicting arguments~\cite{Karlsen2017Trench}. 


\section{Conclusion}
While the emergence of echo chambers can seem to be an unstoppable wave, we must realize that there is hope for a better information ecosystem. We showed that echo chambers are largely a byproduct of recommender systems. As such, what has been manufactured by these systems can likewise be deconstructed by these systems. Social media may not currently live up to its' promise of bringing us closer together and fostering better conversations presently. The future does not have to be this way -- through research and a structured strategy a less polarized world is possible. 

The echo chamber effect is tremendous, as seen in the spread of misinformation amid the COVID-19 pandemic, which has caused many people to lose their lives. One of the most significant abettors to the spread of misinformation are social media platforms and their recommender algorithms that compound the problem of echo chambers. Thus, there is a great need for the social computing community to take the responsibility to develop effective models and tools to help combat the negative effect of echo chambers. Therefore, this survey focuses on a computational perspective to help readers grasp the recent technologies for detecting and preventing echo chambers.

 We detailed the mechanisms that lead to the formation of echo chambers. In summary, the driver behind the formation and growth of echo chambers is the feedback loop between (1)~recommender systems, (2)~psychological biases, namely confirmation bias and cognitive dissonance, and (3)~the homophilic users' networks. We showed that content recommender systems (which are designed to keep the user engaged in social media to watch more ads) are the main reason for the formation of echo chambers. We also showed that the users have the requisite biases to fall into an echo chamber: confirmation bias and cognitive dissonance. Recommender systems utilize these biases to ``personalize'' their recommendations which trap users in echo chambers.

In closing, the echo chamber phenomenon is challenging to tackle because not all stakeholders necessarily want it to be solved. Social media wants users to stay engaged with their platform to show them more personalized ads. Furthermore, it is difficult for people to admit that they live in an echo chamber of users with similar opinions. It is very difficult for one to admit that the other side might hold interesting opinions. However, the way out of echo chambers starts with understanding the mechanisms that lead to the formations of echo chambers and using this knowledge to detect and mitigate echo chambers.

\bigskip \noindent \textbf{Acknowledgement.} We would like to thank Dr. H. Russell Bernard for his insightful input and kind help in this work. He was very generous with his valuable time, and his advice was important for the completion of this work.

\bibliography{ref/mansooreh,ref/faisal,ref/anique,ref/lu,ref/bohan,ref/tyler}


\section*{Appendix}
\appendix

\section{Data Sets}
\label{appendix::data}
In table \ref{tab:datasets}, we list some of the datasets that could be used in any future work related to echo chambers or any related topic, such as political polarization and filter bubbles.

\begin{table*}

\begin{tabular} { 
  | >{\arraybackslash}p{0.30\textwidth} 
  | >{\centering\arraybackslash}p{0.20\textwidth} 
  | >{\centering\arraybackslash}p{0.10\textwidth} 
  | >{\centering\arraybackslash}p{0.10\textwidth} 
  | >{\centering\arraybackslash}p{0.10\textwidth}  | }
        \hline
        \textbf{Dataset Name} &
        \textbf{Number of Samples} &
        \textbf{Time Period} &
        \textbf{Platform} &
        \textbf{Source} \\
        
        \hline \hline
        
        A Long-Term Analysis of Polarization on Twitter - Followers Data &
        140M users &
        2009 - 2016 &
        Twitter &
        \cite{Garimella2017Longterm}
        \\
        
        \hline
        A Long-Term Analysis of Polarization on Twitter - Retweeters Data &
        2B tweets by 679K users & 
        2007 - 2016 & 
        Twitter &
        \cite{Garimella2017Longterm} \\
        
        \hline
        Tweeting From Left to Right &
        150M tweet by 3.8 users &
        2012 - 2014 &
        Twitter & 
        \cite{Barbera2015Tweeting} \\

        \hline
        Segregation and Polarization in Urban Areas & 
        87M tweets by 2.8M users &
        2013 - 2014 &
        Twitter & 
        \cite{Morales2019Segregation}\\
        
        \hline
        Measuring Political Polarization &
        16M tweets by 3M users &
        2013 & 
        Twitter &
        \cite{Morales2015Measuring} \\
        
        \hline
        What is Gab? A Bastion of Free Speech or an Alt-Right Echo Chamber? & 
        22M posts by 336K users & 
        2016 - 2018 & 
        Gab & 
        \cite{Zannettou2018Gab}
        \\
        
        \hline
        Gab Dataset of ``The Echo chamber Effect on Social Media'' & 
        13M post by 165K user & 
        2017 & 
        Gab &
        \cite{Cinelli2021Effect} \\
        
        \hline
        Endogenetic Structure of Filter Bubble in Social Networks &
        1.3M messages &
        2018 &
        Weibo &
        \cite{Min2019Endogenetic} \\
        \hline
        
\end{tabular}

\caption{Echo chamber datasets}
\label{tab:datasets}

\end{table*}

\section{Glossary}
\label{appendix::glossary}
\begin{itemize}
    \item \textbf{Cognitive Dissonance:} refers to an internal contradiction between two opinions, beliefs, or items of knowledge~\cite{Festinger1957ADissonance}.
    
    \item \textbf{Confirmation bias:} is the tendency to seek, interpret, favor, and recall information adhering to preexisting opinions~\cite{nickerson1998confirmation}.

    \item \textbf{Echo chamber:} a network of users in which users only consume opinions that support their pre-existing beliefs and opinions and exclude and discredit other viewpoints..
    
    \item \textbf{Filter bubble:} an environment and especially an online environment in which people are exposed only to opinions and information that conform to their existing beliefs \footnote{\url{https://www.merriam-webster.com/dictionary/filter\%20bubble}}.
    
    \item \textbf{Homophily:} also known as love of the same, is the process by which similar individuals become friends or connected due to their high similarity~\cite{Zafarani2014SocialIntroduction}.
    
    \item \textbf{Political polarization:} is the divergence of political attitudes to ideological extremes\footnote{\url{https://en.wikipedia.org/wiki/Political_polarization}}.
    
    \item \textbf{Selective exposure:} A tendency for people both consciously and unconsciously to seek out material that supports their existing attitudes and opinions and to actively avoid material that challenges their views\footnote{\url{https://www.oxfordreference.com/view/10.1093/oi/authority.20110803100452931}}. 
\end{itemize}

\end{document}